\documentclass[letterpaper, 10 pt, conference]{ieeeconf}  
\usepackage{amssymb}
\usepackage[dvipsnames]{xcolor} 
\usepackage{amsmath}
\usepackage{cancel}
\usepackage{graphics} 
\usepackage{epsfig} 
\usepackage{mathptmx} 
\usepackage{times} 
\usepackage{amsmath} 
\usepackage{amssymb}  
\usepackage{comment}

\usepackage{algorithmic}
\usepackage{algorithm}

\usepackage{lineno}

\IEEEoverridecommandlockouts                              


\usepackage{stackrel,amssymb}
\newcommand{\leftrarrows}{\mathrel{\raise.75ex\hbox{\oalign{%
  $\scriptstyle\leftarrow$\cr
  \vrule width0pt height.5ex$\hfil\scriptstyle\relbar$\cr}}}}
\newcommand{\lrightarrows}{\mathrel{\raise.75ex\hbox{\oalign{%
  $\scriptstyle\relbar$\hfil\cr
  $\scriptstyle\vrule width0pt height.5ex\smash\rightarrow$\cr}}}}
\newcommand{\Rrelbar}{\mathrel{\raise.75ex\hbox{\oalign{%
  $\scriptstyle\relbar$\cr
  \vrule width0pt height.5ex$\scriptstyle\relbar$}}}}

\title{\LARGE \bf
Taming Waves: A Physically-Interpretable Machine Learning\\Framework~for~Realizable~Control~of~Wave~Dynamics
}

\author{Tristan Shah$^{1}$, Feruza Amirkulova$^{2}$, Stas Tiomkin$^{1,*}$\\{\small$^{1}$Computer Engineering Department, \small$^{1}$Mechanical Engineering Department}\\{\small Charles W. Davidson College Of Engineering,
San Jose State University, CA, USA}
\\{\hspace{0.7cm}\small \{tristan.shah, feruza.amirkulova, stas.tiomkin\}@sjsu.edu}
\thanks{$^*$ Corresponding Author}
}

\makeatletter
\let\NAT@parse\undefined
\makeatother
\usepackage{hyperref}

\begin{document}

\pagenumbering{arabic}

\maketitle
\thispagestyle{empty}
\pagestyle{empty}

\begin{abstract}
Controlling systems governed by partial differential equations is an inherently hard problem. Specifically, control of wave dynamics is challenging due to additional physical constraints and intrinsic properties of wave phenomena such as dissipation, attenuation, reflection, and scattering. In this work, we introduce an environment designed for the study of the control of acoustic waves by actuated metamaterial designs. We utilize this environment for the development of a novel machine-learning method, based on deep neural networks, for efficiently learning the dynamics of an acoustic PDE from samples. Our model is fully interpretable and maps physical constraints and intrinsic properties of the real acoustic environment into its latent representation of information. Within our model we use a trainable \textit{perfectly matched layer} to explicitly learn the property of acoustic energy dissipation. Our model can be used to predict and control scattered wave energy. The capabilities of our model are demonstrated on an important problem in acoustics, which is the minimization of total scattered energy. Furthermore, we show that the prediction of scattered energy by our model generalizes in time and can be extended to long time horizons. We make our code repository publicly available.
\end{abstract}


\section{Introduction}\label{sec:Introduction}

\begin{figure*}[t!]
    \includegraphics[width=\textwidth]{"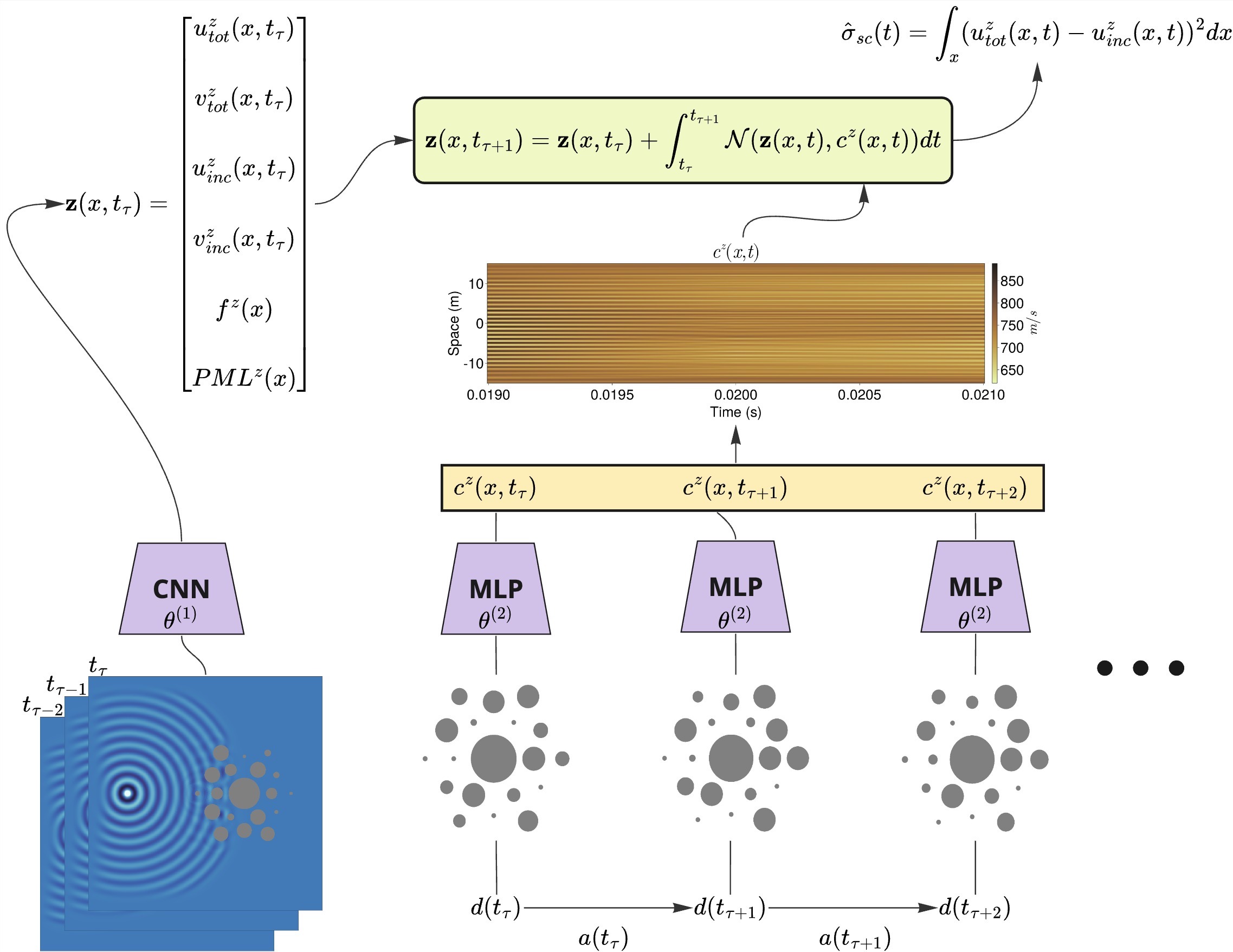"}
    \caption{Overall pipeline of our algorithm which shows the flattening of 2D partial information into the initial conditions of a 1D wave simulation. Information supplied to the wave encoder comes from the stack of "wave images" shown in blue at timesteps: $t_{\tau-2}, t_{\tau-1}, t_\tau$ is input to the CNN. The source of acoustic energy is a point which appears in white. Concentric rings emanate from this point and scatter upon the design shown in grey. Information about the sequence of designs is input to the design encoder and used to construct a speed of sound function $c^z(x, t)$ which influences the latent solution. The relevant information is combined and used to propagate the latent dynamics $\mathbf{z}(x, t_\tau)$. Finally, scattered energy is computed in latent space and used to approximate scattered energy in the real environment.
    }
    \label{fig:model}
\end{figure*}

Partial Differential Equations (PDE)s are useful mathematical tools for describing how physical systems behave as a continuum. Long established PDEs such as the classical wave equation, Navier-Stokes equation, heat equation, and the Schrödinger equation govern systems which are the foundation of our understanding of physics. When combined together and with other observations of physical systems they can be used to describe complex phenomenon such as wild fires \cite{javaloyes_2023}, the global climate \cite{schneider_1975}, or spread of disease \cite{majid_2021}.

Control over systems governed by PDEs usually requires the existence of an analytical solution which is dependent on control parameters to influence the system in a desirable direction \cite{shah_2021, cominelli_2022}. With the rise of powerful sample based learning methods it is now possible to influence PDEs without deriving a solution \cite{bieker_2020, werner_2023}, unlocking control of a much larger suite of problems in physics than was previously accessible. However, employing algorithms from machine learning to solve problems in PDE control is not always computationally tractable. PDEs are difficult to model with neural networks (NN) due to their infinite-dimensional tensor (function valued) states and sensitive dynamics \cite{pan_2018}. While machine learning (ML) methods have been shown to be very successful at controlling simple scenarios, it is important to search for ways to reduce the complexity of this type of problem for the development of truly scalable methods. 

Establishing control over the acoustic wave equation is important because it provides pathways to developing new technologies. These technologies range from acoustic cloaking for airborne and aquatic stealth, wave steering devices for seismic wave manipulation, to 'super-focusing' devices for high-resolution ultrasound imaging and surgery. There exist some known analytical solutions to this PDE \cite{norris_2018, cominelli_2022} which have been used to solve these problems. However, in our work, we pursue an optimal control method that leverages ML. 

In our research, we design an algorithm for prediction and control of acoustic PDEs that could eventually be realized by a physical device. We allow our model to observe partial information from its environment which could realistically be collected by sensors. Furthermore, we consider a control mechanism capable of being constructed and deployed. We also make advancements in interpretability of dynamical models learnt from samples. We design our model with a latent space structured by the wave equation, which allows us to visualize and draw meaning from its state. We use the interpretable structure of the wave equation to introduce a novel and trainable energy dissipation layer derived from domain knowledge in acoustics. Overall our work represents an advancement in the field of PDE control which will enable future works to build more interpretable and robust artificial intelligence.

\subsubsection{Contributions}

The results of our work can be distilled into three major contributions:
\begin{itemize}
    \item Environment for studying the behavior of waves in free field domains in response to a controller.
    \item Algorithm for learning the dynamics of \textit{scattered energy} in response to a control signal.
    \item A novel trainable dissipation layer which can learn to explicitly remove energy from latent space.
\end{itemize}

We elaborate on our method in Section~Method. 

\section{Prior Work}\label{sec:Prior Work}

In recent years, the control of systems governed by PDEs has become an increasingly active topic in ML \cite{cominelli_2022, shah_2021, werner_2023, bieker_2020, wu_2022}. Research in this field falls into two categories. One can be thought of as "parameter optimization" of a PDE solution. It requires a known solution to a PDE which can be used to form an optimization objective based on some parameters. An algorithm such as gradient-based optimization (GBO) \cite{cominelli_2022} or Reinforcement Learning (RL) \cite{shah_2021} is used to optimize the design parameters. However, the requirement for a PDE solution limits its applicability. The other category is dynamical control of PDEs. This category does not require knowledge of a PDE solution. The method takes a PDE, incorporates a control signal into its dynamics, and uses methods like Reinforcement Learning (RL) \cite{werner_2023} or Model Predictive Control (MPC) \cite{bieker_2020} to select appropriate control actions. The limitation of this work is its use of an RNN to learn PDE dynamics. RNNs lack interpretability \cite{rahman_2021} and tend to overfit \cite{zaremba_2015}. In the following text, we elaborate on these two categories and overview the distinguishing features of the current work, which are explained in detail in Section~Method.

Parameter optimization has been used \cite{cominelli_2022} to design arbitrarily shaped acoustic cloaks. In their work, Cominelli et al. formulate an objective function using a known solution to the wave equation. Their objective is dependent on material properties in a region of a computational domain and predicts scattered wave energy. Material properties are optimized using GBO such that scattered energy is minimized, resulting in an acoustic cloak. Similarly, \cite{shah_2021} used the similar approach to design a cloaking device by optimizing the position and radii of cylindrical scatterer configurations. Scattered wave energy is minimized by RL agents which explore the design parameter space for optimal configurations. Optimization of design parameters through known PDE solutions is a robust and precise method but is limited by the existence of such solutions. In many cases there are no analytical formulas to solve PDEs making parameter optimization infeasible in many important scenarios.

Recently, a formulation that incorporates the dynamics of a PDE into an optimal control problem has been explored \cite{werner_2023}. The method in \cite{werner_2023} has the advantage of not requiring a known solution in order to be applicable; widening the range of problems where it can be employed. The idea proposed in \cite{werner_2023} is the numerical integration of PDE dynamics while a control signal is applied. The authors in \cite{werner_2023} consider a particular PDE, the 1D Kuramoto-Sivashinsky (KS) PDE, where control $\phi: \Omega\times [0, T] \rightarrow\mathbb{R}^m$ is an additive force applied to system variable $u: \Omega\times [0, T]\rightarrow\mathbb{R}^n$. In their work $\phi$ is fixed over a time interval and is updated between subsequent intervals. Equation~\eqref{eq:optimal-control-integration} shows integration of the PDE over an interval where $\phi$ is being applied concurrently with the dynamics resulting in a next state $u_{\tau+1}$. This formulation is interesting because it unifies the dynamics that we wish to control with the application of a control signal. Methods from optimal control such as Reinforcement Learning (RL) or Model Predictive Control (MPC) can be applied to manage the control signal for optimization of any particular objective function given in Equation~\eqref{eq:oc-objective}. An important feature of their work is the efficient manner in which $\phi$ is constructed. The authors in \cite{werner_2023} consider $\phi$ as a superposition of four Gaussian distributions with fixed variance and position. The effect of the controller on $\phi$ is to modulate the intensity of each Gaussian through the $\mathbf{a}_\tau[i]$ term. This procedure significantly reduces the dimension of the control input from $\Omega$ to $\mathbb{R}^4$. Reduced dimensional construction of a control signal is a highly relevant technique to our work as we wish to control waves in a two spacial dimensions. Generating control over a high resolution finite element grid may be computationally intractable without a parameterized method. This work provides useful formalism for framing a PDE as an optimal control problem but is limited by the simplicity of the scenario it considers. 

Other work \cite{bieker_2020} also considers dynamical control over the Navier-Stokes equation. Their research focuses on designing a physically realizable control algorithm. In their work, Bieker et al. trained a system of deep neural networks to predict lift and drag coefficients on a configuration of rotating cylinders over a time horizon. A recurrent neural network (RNN) was used to approximate these quantities by taking in the state of the simulator and a sequence of control inputs. Bieker et al. show that their RNN dynamics model is able to make accurate predictions and performs optimal control with MPC. There are two particular contributions made in this work which are relevant to our research. Firstly, the model takes in information gathered from physically realizable sensors rather than the full state of the fluid simulation. This practice makes the model much more applicable in real world scenarios when limited data is available. Secondly, the effect of the control input on the PDE is locally sparse. In other words, the authors do not consider the ability to influence the PDE at every point in the computational domain but only at the interface between the cylinders and the ambient fluid. The effect of sparse control is indirect compared to the alternative but more realistically achievable with a physical device. This work is relevant to our research in its proposal of parameterized construction of a control signal but is limited to a relatively simple system which occurs in 1D. Additionally, the dynamics model is not interpretable and does not incorporate domain knowledge as helpful structure.
\section{Current Work}\label{sec:Current Work}
\subsection{Preliminaries}\label{subsec:Preliminaries}


\subsubsection{Acoustic Metamaterials}
Metamaterials are engineered materials that exert an influence over acoustic waves in the extraordinary form of sound light cloaking, super-focusing,   negative refraction, negative mass density, and negative Poisson's ratio to name a few. 
The realization of novel acoustic materials with enhanced capabilities provides the potential expanded functionality over a 
range of applications including cloaking \cite{Norris08b, Amirkulova2020}, wave steering, and  focusing. 
In this work, 
we design metastructures
focusing on sound direction control such as cloaking by suppressing the scattered wave energy.

\subsubsection{Acoustic Multiple Scattering}
Multiple scattering \cite{Amirkulova2020} from fluid media with dissimilar material properties
is of great importance to various applications in
acoustics and elastic wave propagation. For instance, biomedical
ultrasound and non-distractive testing
make use of acoustic field scattering
for localization, imaging, and object detection. In this work, we consider a 2D planar motion and  multiple scattering by cylindrical structures in acoustic media. For demonstration purposes, we consider
a cluster of fluid cylinders located inside another fluid medium.
A 
realization of such composite fluid media can be achieved by  using effective medium theory \cite{dt:06}. 
The governing equation for sound waves 
is  a wave equation:
\begin{equation}
    \ddot{u} = c^2\nabla^2 u
\end{equation}
where $\nabla^2 $ is the Laplacian, $t$ is time, $u$ is  the 
physical quantity such as 
pressure or velocity potential, and $c$ is the sound speed in a fluid medium.
The boundary conditions at the interface around each scatterer are the continuity of normal components of the particle velocity and normal stress components.
\subsection{Free Field Scattered Wave Environment}\label{sec:Environment}

The environment we introduce simulates interaction between acoustic waves and metamaterial designs actuated by a controller. It serves as a platform for studying control of wave dynamics in open space. A key feature of our environment is its capability to simulate a "free field" infinite spacial domain. Many problems in acoustics rely on the assumption that the space in which waves propagate is infinite \cite{johnson_2021}; no reflections occur along the boundary of the domain. We make our environment conveniently accessible to the ML community through an interface in the Julia Programming Language. Our goal is to promote the use of problems in acoustics as a testing ground for PDE control algorithms. Our environment and its implementation offer the technical tools required to further explore this area and establish stronger connections between acoustics and ML. The structure of our environment and important features are described in the following text.

\subsubsection{Excitation Force} Our environment differs from environments in classical control because it has a time-dependent excitation which continously adds energy to the system. Moreover, the excitation is not directly controllable by the agent we consider in this work. Sources in our environment can have an arbitrary shape defined by a function $f(\mathbf{x})$ in Equations~(\ref{eq:2d-wave-2}, \ref{eq:2d-wave-3}) which oscillate with frequency $\omega$. This flexible definition of the source function enables our environment to simulate different scenarios such as plane waves or point sources. It is important to note that if an excitation exists in a closed environment the energy level will constantly increase. In order to study how acoustic waves interact with a non-linearity in isolation from energy buildup, dissipation must be incorporated \cite{johnson_2021}. Without such dissipation, energy will reflect on the boundaries of the domain multiple times and interfere with scattered energy in the area of interest. 

\subsubsection{Perfectly Matched Layer} One of the most important distinguishing features of our environment is its free field capability. We simulate infinite space in two dimensions using a Perfectly Matched Layer (PML). A PML is an artificial material placed around the boundary region of a computational domain that exponentially dampens waves as they propagate \cite{prigge_2016}. Dampening waves causes their energy to approach zero and never return to the domain. The main idea behind a PML is to apply complex coordinate stretching in the frequency domain as waves enter the PML region. The transformation shown in Equation~\ref{eq:pml_transform} is applied to a wave equation in 1D and scales spacial derivatives to become smaller farther into the PML. This transformation can be applied to wave equations in any dimension as long as the coordinate stretching procedure is applied to each spacial partial derivative. The application of the PML transformation to the two-dimensional wave equation results in fields $\psi_x$, $\psi_y$, and $\gamma$ which dissipate energy along the $x$ and $y$ axes and corners respectively. The evolution of these fields in time is determined by Equations~(\ref{eq:2d-wave-4}-\ref{eq:2d-wave-6}). The resulting dissipative fields influence the solution variables $u$, $v_x$, and $v_y$ in Equations~(\ref{eq:2d-wave-1}-\ref{eq:2d-wave-3}). Functions $\sigma_x$ and $\sigma_y$ determine the shape of the PML in their respective axes in 2D. These functions have a value of 0.0 everywhere in the domain and slope up towards 1.0 inside the boundary region:

\begin{equation}\label{eq:pml_transform}
    \frac{\partial}{\partial x} \rightarrow \frac{\partial}{\partial x} \cdot \frac{1}{1 + i\frac{\sigma_x}{\omega}}.
\end{equation}

\subsubsection{Design Representation} The particular type of metamaterial designs we implement in our environment are fluid scatterers. Specifically, we consider cylindrical scatterers with radial symmetry; though in principle other shapes are achievable in this framework. The scatterers influence the behavior of an incident wave by reflecting a portion of its energy. This phenomenon occurs through a difference in material properties at the interface between the scatterer and the ambient domain. A benefit of using such scatterers is their influence on the wave environment is physically plausible and could be achieved by real devices. This method of actuating the environment is spatially sparse because the control cannot affect the wave at every point in the field. Sparse control in our work is in contrast to other works which consider an agent with the ability to affect the PDE solution at every point in space \cite{werner_2023}. Not only is non-sparse control usually not realistic, but it is also extremely computationally complex to optimize a controller to output function valued actions at the same scale as the computational domain. In our framework, a design can be expressed by an indicator function defined by Equation~\eqref{eq:wave-speed} which assigns the appropriate speed of sound to every point in the domain at time $t$

\begin{equation}\label{eq:wave-speed}
    c_\tau(\mathbf{x}, t) =
    \begin{cases}
        1032\;m/s & \mathbf{x} \in d_\tau(t), \\
        1531\;m/s & else.
    \end{cases}
\end{equation}

\subsubsection{Control by Agent} Actions in this framework represent an agent's control over the metamaterial design. For the particular design that we consider in our experiments, the agent actuates its control by adjusting the radii of each scatterer. This adjustment is selected at a time step and applied over the time interval at a rate of $500m/s$. We use linear interpolation between the parameters (radii) of designs to represent a smooth transition between the design before and after the application of the action. The indicator function for material properties can be evaluated at any time in this interpolation to obtain the information necessary to propagate the wave simulator. In our case, we consider the ambient domain as water with a speed of sound of $1531m/s$ and the scatterers as a fluid with three times the speed of sound in the air of $1032m/s$. These values were chosen for the universal significance of water, however our simulator can support any nonzero value of sound speed. The propagation of waves in our simulator involves numerical integration of the PML-transformed Equation~\eqref{eq:2d-wave-1} with $c_\tau$ as the indicator function. Once the time of the simulation has reached the end of the interpolation, a new action is applied to the design which forms the next interpolation interval $c_{\tau+1}$.
\begin{align}
 \dot{u}      &=c_\tau^2(\partial_xv_x + \partial_yv_y) + \psi_x + \psi_y \nonumber\\
 &\qquad\qquad\qquad\qquad- (\sigma_x + \sigma_y) u - \gamma,\label{eq:2d-wave-1}\\
        \dot{v}_x        &=\partial_x(u + f\cdot\sin(2\pi\omega t)) - \sigma_x v_x, \label{eq:2d-wave-2}\\
        \dot{v}_y        &=\partial_y(u + f\cdot\sin(2\pi\omega t)) - \sigma_y v_y, \label{eq:2d-wave-3}\\
        \dot{\psi}_x     &=c_\tau^2\sigma_x\partial_yv_y, \label{eq:2d-wave-4}\\
        \dot{\psi}_y     &=c_\tau^2\sigma_y\partial_xv_x, \label{eq:2d-wave-5}\\
        \dot{\gamma}     &=\sigma_x\sigma_y u. \label{eq:2d-wave-6}
\end{align}
The full derivation of these equations appears in supplementary materials. 

\subsubsection{Real Dynamics}
The dynamics of the system variables describing displacement $u = u(\mathbf{x}, t)$ and velocity $v_x = v_x(\mathbf{x}, t)$ and $v_y = v_y(\mathbf{x}, t)$ are defined by Equations~(\ref{eq:2d-wave-1}, \ref{eq:2d-wave-2}, \ref{eq:2d-wave-3}). The displacement field is influenced by the divergence of the velocity field weighted by the speed of sound function $c_\tau$. Here, the operators $\partial_x$ and $\partial_y$ represent partial derivatives $\frac{\partial}{\partial x}$ and $\frac{\partial}{\partial y}$. Dissipation from the PML is incorporated through the inclusion of fields $\psi_x=\psi_x(\mathbf{x}, t)$, $\psi_y=\psi_y(\mathbf{x}, t)$, and $\gamma(\mathbf{x}, t)$. Similarly, the velocity fields are also updated with the appropriate partial derivatives of the displacement field. The energy source is incorporated by the addition of a forcing term $f = f(\mathbf{x})$ multiplied by a sinusoidal function on time. Together the full state of this PDE system can be represented by Equation~\eqref{eq:rd-state}. To ensure the stability of numerical integration of this system a fourth order Runge-Kutta scheme is used:
\begin{equation}\label{eq:rd-state}
     \mathbf{w} = \begin{bmatrix}u & v_x & v_y & \psi_x & \psi_y & \gamma\end{bmatrix}^\mathsf{T}.
\end{equation}

\subsubsection{Partial State Observation} In our work we develop an environment which simulates realistic interaction between a controller and acoustic waves. In such a setting full observability of the environment is not achievable. We consider information which can be obtained through sensors such as readings of the total pressure field at points in the simulation domain. The environment state representations that we use are in the form of reduced resolution images ($128\times128$) from the full discretized finite element grid ($700\times 700$) which are derived from the displacement of the total wave $u(\mathbf{x}, t)$ over the domain evaluated at times $t_{\tau-2}$, $t_{\tau-1}$, and $t_\tau$. We do not include images of the underlying velocity components $v_x$ and $v_y$ because they would not be directly observable in a real environment. In our case we used the reduced resolution displacement images as input to our algorithm but in principle an arbitrary sensor reading can be simulated to provide more realistic observation.

\subsubsection{Training Signal} Various quantities can be computed from the state of our simulator and used as reward or cost signals for a control algorithm. One possibility is to control the energy in the scattered acoustic field. Scattered waves are acoustic fields which result from the interaction between a wave and an obstacle. This field can be obtained by Equation~\eqref{eq:seperation}. In this definition the total field $u$ is the full waveform which accounts for an incident wave plus the scattering it produces through interaction with a boundary condition. The incident field $u_{inc}$ is a solution which accounts for the behavior of waves emanating from the source $f(\mathbf{x})$ without the presence of a design that produces scattering. In other words $u_{inc}$ assumes the entire domain to have a constant rate of sound propagation. These fields are represented by two separate wave dynamics which share the same grid size, ambient speed of sound, initial conditions, and source. To obtain the scattered energy field we subtract the two solutions. The scattered field is important because this quantity is directly dependent on the behavior of the metamaterial design and by extension the controller. The particular objective we demonstrate in this paper is the reduction in total scattered energy $\sigma_{sc}$ which is defined by the integral: 
\begin{equation}\label{eq:sigma}
    \sigma_{sc}(t) = \int_{\mathbf{x}\in\Omega}u_{sc}(\mathbf{x}, t)^2 d\Omega
\end{equation}
where
\begin{equation}\label{eq:seperation}
    u_{sc} = u - u_{inc}.
\end{equation}

\subsection{Algorithm}\label{subsec:Method}

We have developed a model that predicts scattered wave energy $\sigma_{sc}(t)$ as agent's actions are applied to a design in our simulator. The full schematic of the model is shown in Fig~\ref{fig:model} and details information pathways from the wave and design sequence to the latent dynamics and decoder. Our model has several features we will highlight in this section including its latent dynamics, interpretability, and ability to make predictions with partial information. 

\subsubsection{Reduced-Dimensionality Latent Wave Dynamics} The latent dynamics (LD) component of our model shown in the integration of $\mathbf{z}$ in Fig~\ref{fig:model} is a compressed, lower dimensional analog to real dynamics present in our environment. LD share physical correspondence with real dynamics (RD) in the form of learnable initial conditions, excitation force, material properties, and energy dissipation. The key distinction of LD is its lower dimensional state which is 1D. To accommodate the compression of the RD into a lower dimensional space while maintaining as much information as possible we assume that the LD must occur in a physically large space.

\subsubsection{Wave Encoder}

In our model partial state information of the wave simulation in 2D is processed by the wave encoder (CNN) and transformed into full information initial conditions and parameters of 1D LD. We simulate two separate LD (total and incident) and subtract their displacement fields to obtain a scattered energy field through Equation~\eqref{eq:seperation}. To simulate both fields, the initial conditions of the LD are output from the wave encoder $(u^z_{tot}, v^z_{tot}, u^z_{inc}, v^z_{inc})$ which are discretized functions over 1D space. The wave encoder also produces a learnable excitation force function $f^z$ which oscillates at a known frequency of the source in RD. This force continuously adds energy to the LD. Lastly, the wave encoder learns the shape function of the PML in latent space, allowing it to control how much energy is dissipated.

\subsubsection{Design Encoder} Nonuniform distribution of material properties in space influences the RD and results in scattering of acoustic waves by affecting the rate at which sound propagates: $c$. The same effect can be achieved in LD through a 1D function which defines material properties in the latent space. It is a natural choice to use information from the state of the metamaterial design to construct latent material properties $c^z(x, t)$. We use a design encoder (MLP) to map the parameters (radii of each scatterer in the configuration) of the sequence of design states to corresponding discretized functions in 1D. Linear interpolation between the functions produces a continuous latent speed of sound function $c^z(x, t)$ which is defined over the same time interval as the actions are applied in RD. This interpolation between latent representations of designs can be visualized in Fig~\ref{fig:model}. The latent speed of sound $c^z$ function is applied only to the latent total field $u^z_{tot}$ while $u^z_{inc}$ has a constant rate of sound propagation which is assumed to be the same as the ambient speed of sound in RD.
                      
\subsubsection{Sinusoidal Embedding} We use a linear superposition of $\sin$ waves to construct continuous latent fields with a fixed number of frequencies $\sum_{n=1}^N b_n\sin(\frac{n\pi x}{L})$. Intuition from this design choice comes from the general solution to the wave equation in 1D in which initial conditions can be described in this manner. Sinusoidal embedding is applied to the outputs of both wave and design encoders. Instead of producing the full latent fields as their raw outputs, encoders learn a vector of frequency coefficients $b_n^{(i)}$ for each field. The sinusoidal basis functions have fixed endpoints of zero at the boundaries of the latent domain.

\subsubsection{Forward Pass}
The initial conditions excitation force, and PML functions are obtained from the wave encoder $\mathbf{z}(t_\tau)$ and the interpolation of material properties function $c^z(x, t)$ from the design encoder. Conditions are propagated by the LD to obtain a full solution $\mathbf{z}(\cdot)$. LD can be solved with and without a PML to test the necessity for dissipation of energy when predicting $\sigma_{sc}$ in free field. To our knowledge our work is the first to ever use a PML to dissipate energy in the latent dynamics of a neural network. The LD solution is obtained by propagating wave dynamics over the time interval that actions are applied and $\sigma_{sc}(t)$ is measured in RD. Similarly to the RD a Runge-Kutta integrator is used to integrate LD. The predicted scattered energy signal is computed from $\mathbf{z}(\cdot)$ by obtaining $u^z_{sc}$ through Equation~\eqref{eq:seperation} and integrating the energy over 1D space $\hat{\sigma}(t)_{sc} = \int_x u^z_{sc}(x, t) ^ 2 dx$. Prediction error between true and predicted signals is measured through mean squared error loss $\int_t(\sigma_{sc}(t)-\hat{\sigma}_{sc}(t))^2dt$. Additionally, $\int_t(\sigma_{tot}(t)-\hat{\sigma}_{tot}(t))^2dt$ and $\int_t(\sigma_{inc}(t)-\hat{\sigma}_{inc}(t))^2dt$ are minimized as auxiliary constraints.

\subsubsection{Backwards Pass} Obtaining gradients of the latent solution with respect to the prediction error relies on auto-differentiation (AD). These gradients can be computed using reverse mode differentiation in deep learning frameworks \cite{Flux.jl-2018}. However, in order to train the wave encoder and the design encoder, gradients w.r.t the initial latent conditions, parameters $\mathbf{z}(x, t_\tau)$, and latent speed of sound $c^z(x, t)$ must be computed since these are the outputs of the encoders. Standard AD methods are not best suited to take gradients through the integration of a PDE. Instead we have used the adjoint sensitivity method \cite{chen_2019} which is faster and more memory-efficient. Adjoint sensitivity works by treating backpropagation as the reverse integration of a differential equation where the solution variable itself is a gradient. Reverse integration of the gradient w.r.t. each time step of LD to the initial condition allows us to obtain the information necessary for updating the parameters of the encoders. 

\subsubsection{Model Predictive Control} We use model predictive control (MPC) to demonstrate our model's ability to influence the quantity that it predicts. We optimize a sequence of actions under the objective in Equation~\eqref{eq:oc-objective}. In our case, the cost Equation~\eqref{eq:ell-definition} is summed over the number of actions in the sequence $\tau = 1,\dots,n$ where $\beta$ is a Lagrange Multiplier that penalizes large actions. Random shooting with $256$ shots and a prediction horizon of $10$ is used to generate and select an action sequence that is optimal under the given objective. The first action of the optimal sequence is applied to the RD and this process repeats. We demonstrate the performance of our model in Section Experiments.

Equations~\eqref{eq:oc-objective}-\eqref{eq:differential-operator} are adapted from the work of \cite{werner_2023} and serve as a formulation for framing PDE control as an optimal control problem. We minimize the following functional:

\begin{equation}
    \min_a \; J(\mathbf{w}, a) = \min_a\sum_{\tau=1}^{T}\ell(\mathbf{w}_\tau, a_\tau)\label{eq:oc-objective},
\end{equation}
with 
\begin{align}
    \mathbf{w}_{\tau + 1}= &\mathbf{w}_\tau +  \int_{t_\tau}^{t_{\tau + 1}}\mathcal{N}(\mathbf{w}(x, t), a(t))dt \label{eq:optimal-control-integration}
\end{align}
which is a discrete time version of the PDE in continuous time \cite{werner_2023}:
\begin{align}
    \frac{d\mathbf{w}}{dt} = \mathcal{N}(\mathbf{w}(x,t), a(t)),\label{eq:differential-operator}
\end{align}
where $\mathbf{w}$ given by Equation~\eqref{eq:rd-state} is a vector of fields describing the state of the RD system. The sequence of actions is represented as the time dependent function $a(t)$ that changes the state of the design $d(t)$ over a time interval. In the above equations $\mathcal{N}$ represents a particular PDE operator. In our case this operator is the PML-transformed first order classical wave equation system given in Equations~(\ref{eq:2d-wave-1}-\ref{eq:2d-wave-6}) which acts upon the fields of $\mathbf{w}$. Minimization in Equation~\eqref{eq:oc-objective} occurs over control trajectories, $a(t)$ which are obtained through interpreting a design trajectory $d(t)$ as an interpolation of nonlinear speed of sound functions $c(\mathbf{x}, t)$ in RD. MPC involves optimizing a control trajectory under objective Equation~\eqref{eq:oc-objective}, where $\ell$ is defined Equation~\eqref{eq:ell-definition}, then applying the first action in the sequence $a(t_\tau)$ to RD. The prediction of $\hat\sigma_{sc}(t)$ occurs by representing $\mathbf{w}$ in a learned lower dimensional state $\mathbf{z}$. This state is propagated with influence from predicted latent speed of sound function $c^z(x, t)$ to obtain the full solution $\mathbf{z}(\mathbf{x}, t)$
\begin{equation}\label{eq:ell-definition}
    \ell(\mathbf{w}_\tau, a_\tau)=\int_{t_\tau}^{t_{\tau+1}}\hat\sigma_{sc}(t) dt +\beta\lVert a(t_\tau)\rVert_2^2.
\end{equation}

\section{Experiments}\label{sec:Experiments}
In this section we demonstrate the capabilities of our model on the task of prediction and control of total scattered energy. Several questions are explored and answered:

\begin{enumerate}
    \item Can our model generalize to long planning horizons in test time when it is trained only on a short horizon? 
    \item Is it important to dissipate energy in latent space with a trainable PML layer?
    \item Will a trained model of the wave dynamics be able to control $\sigma_{sc}(t)$?
\end{enumerate}

\subsubsection{Prediction} 
\begin{figure}[t!]
    \centering
    \includegraphics[width=\linewidth]{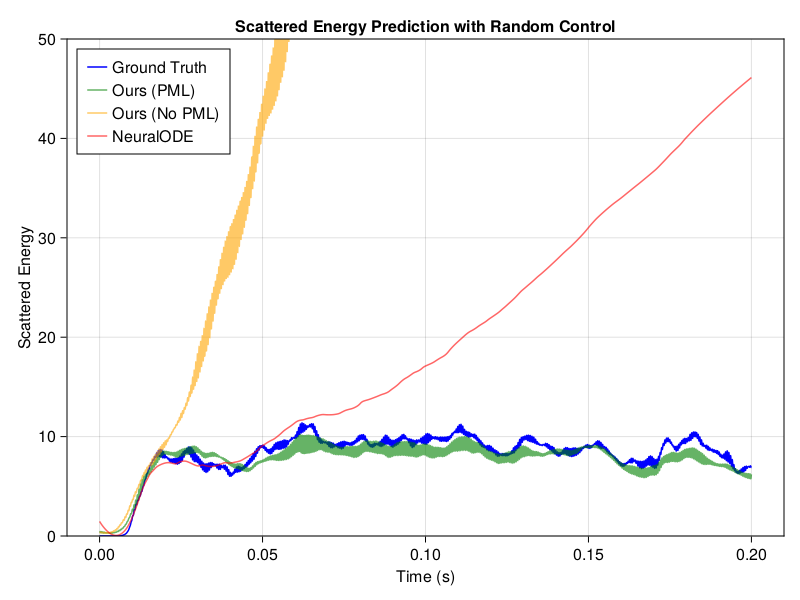}
   \vspace{-0.2in}
    \caption{Demonstration of our models capability to predict $\sigma_{sc}(t)$ over an entire episode from partial information and a sequence of $200$ actions taken in the environment compared against NeuralODE. These curves represent $20000$ integration timesteps of the dynamics. Prediction capabilities of our model with energy dissipation (blue), without (orange), and NeuralODE (red) are compared against true scattered energy (blue).
    \vspace{-0.26in}
    }
    \label{fig:long-prediction}
\end{figure}

The results of predicting $200$ actions ($0.20$s) with a model trained on a horizon of $20$ actions are shown in Fig~\ref{fig:long-prediction}. We run our latent dynamics with and without a PML in the LD integration process in Fig~\ref{fig:model}. It is clear that the PML model is capable of predicting macro level trends in $\sigma_{sc}(t)$ over the entire episode of $20000$ integration steps despite not being explicitly trained to do so; thereby demonstrating qualities of temporal generalization. Without a PML the model fails to successfully predict for longer time horizons than it was exposed to in training. A buildup of energy from the source $f^z$ occurs in latent space due to its inability to leave the closed system. Dissipation of energy in LD allows for stable steady state behavior which can be extended much longer than its training horizon. The SOTA baseline we compare our method to is NeuralODE, a general purpose deep neural network based ordinary differential equations solver. As shown in Fig~\ref{fig:long-prediction} NeuralODE does not immediately diverge from the ground truth signal but tracks it for approximately $75$ actions ($0.075s$). After extending the prediction horizon of NeuralODE beyond $75$ actions it begins to diverge. We hypothesize that this divergence is due to NeuralODEs inability to explicitly dissipate latent energy. Implicitly it must learn some degree of energy dissipation through it's trainable dynamics, however, not enough to make long term stable predictions. Qualitatively, the predictions generated by NeuralODE are rather simplistic. It generates a signal which generally follows the trend of ground truth scattering for sub $75$ action horizons but it fails to capture high frequency oscillations in $\sigma_{sc}$. The predictions generated by our method appear as thicker lines. This is due to the underlying wave dynamics in the latent space of our model which are able to simulate low and high frequency waves.

\begin{figure}
    \centering
    \includegraphics[width=\linewidth]{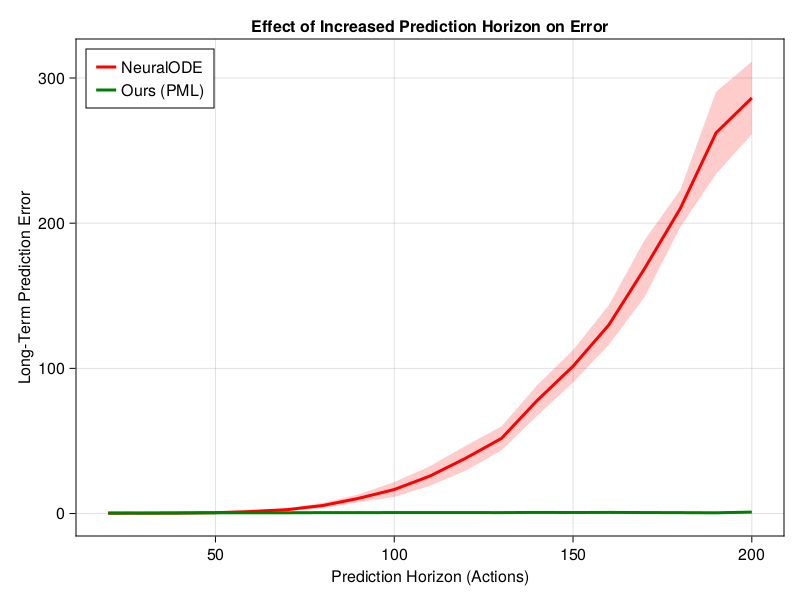}
    \vspace{-0.1in} 
    \caption{Evaluation of our model's prediction error as a function of prediction horizon compared to NeuralODE. Mean squared error is evaluated at horizons from $20$ to $200$ in intervals of $10$. Each evaluation is conducted on a randomly sampled batch of $32$ data points unseen during training.
    \vspace{-0.1in}}
    \label{fig:loss}
\end{figure}
We investigate how the performance of our model scales with increased horizon length compared to NeuralODE in Fig~\ref{fig:loss}. Starting at the training horizon of length $20$ ($0.02s$), the mean squared error loss distribution is visualized using a line plot which shows the mean and one standard deviation. Prediction horizon length is incremented by $10$ actions up to the entire episode length of $200$ ($0.20s$). It can be observed that the mean and variance of loss measured with the PML model is negligible. The error of NeuralODE's predictions is steady at first until approximately $75$ actions beyond which it rapidly degrades. This comparison shows that our model is significantly superior to NeuralODE in long term generalization.

\subsubsection{Effect of PML}
\begin{figure}[t!]
    \centering
    \includegraphics[width=\linewidth]{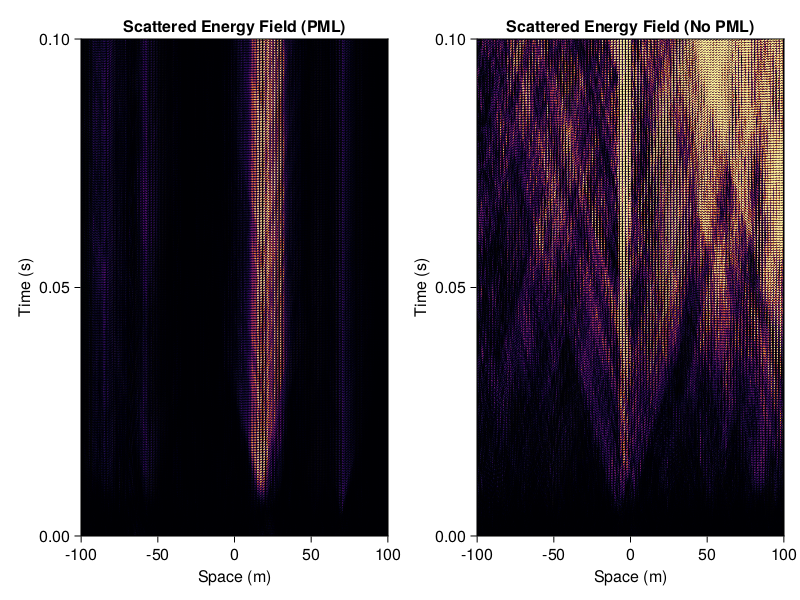}
    \caption{Comparison of the scattered energy fields of our model with and without a latent PML. The field is visualized over $100$ actions ($0.1s$).}
    \label{fig:latent}
\end{figure}

Energy in latent space $u^z_{sc}(x, t)^2$ is visualized in Fig~\ref{fig:latent}. Our model with a trainable PML shown in Fig~\ref{fig:latent} (left) demonstrates that the scattered energy field naturally segments into one major zone of activity and three additional zones which appear very faintly. The PML is learned such that energy dissipates on either side of each zone and that energy from neighboring zones does not build up and interfere with one another. This enables steady energy levels which result in the stable predictions in Fig~\ref{fig:long-prediction} and Fig~\ref{fig:loss}. Our model without a PML (right) shows clear signs of energy buildup. The latent space is unable to be divided into zones and constructive interference causes rapid accumulation of energy. It is clear this model would not be capable of making stable predictions without the ability to dissipate excess energy.

\subsubsection{Control} 
\begin{figure}[t!]
    \centering
    \includegraphics[width=\linewidth]{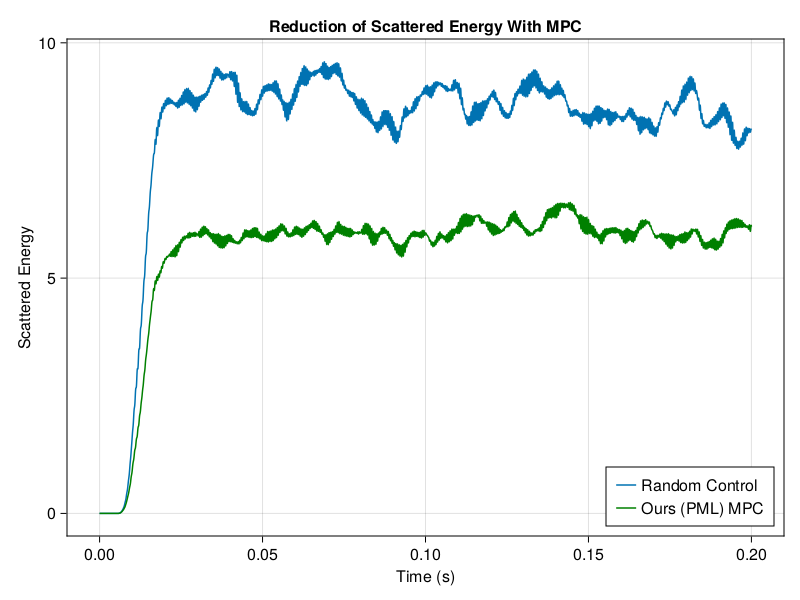}
\vspace{-0.3in}
    \caption{Comparison of $\sigma_{sc}$ produced in the environment when taking random actions (blue) versus actions selected with random shooting MPC (green) with a planning horizon of $10$ actions. Each curve is an average of 6 episodes run with their respective policies.\vspace{-0.2in}}
    \label{fig:mpc-control}
\end{figure}

We used our model with a trainable PML to control $\sigma_{sc}(t)$ in the real environment due to our observation of its generality in the previous result. An MPC agent uses our model's prediction of $\hat\sigma_{sc}(t)$ produced by the LD in Fig~\ref{fig:model} to inform the cost function in Equation~\eqref{eq:ell-definition}. Optimal actions selected by way of MPC and applied to the real environment. We show a significant reduction (approximately $30\%$) in mean $\sigma_{sc}(t)$ over the episodes generated with MPC compared to those with random control in Fig~\ref{fig:mpc-control}. The results of this experiment show that our model's predictions are accurate enough to be used for control. With a more advanced control algorithm the performance could be further improved and possibly result in an effective acoustic cloaking device.


\section{Conclusion}\label{sec:Conclusion}

We have developed an environment and algorithm which can serve as a basis for ML research in continuous dynamical systems governed by wave equations. Our environment is capable of simulating designs which can be manipulated in continuous time and influence complex PDE systems with physically realizable mechanisms. The novel algorithm we introduce operates on partial information from the environment and incorporates domain knowledge from acoustics to structure a latent representation of system dynamics. Our model is interpretable through physical properties, therefore, a level of meaning can be assigned to the evolution of its LD. The interpretable structure of the latent space we are able to apply a novel and trainable PML layer which actively dissipates energy. We discover that energy buildup in latent space is an important attribute and must be dissipated to allow the latent dynamics to maintain stability in the steady state. To our knowledge, a PML has not been incorporated into neural networks before. Furthermore, our work is the first to control of acoustic waves in the time domain with ML.
\section{Acknowledgments}
The authors thank the graduate students at the Computational Intelligence Control and Information Laboratory at Charles W. Davidson College of Engineering SJSU, and especially, Megha Jain and Ezgi Kaya for the discussions and help with proofreading. This work was supported by the Research and Innovation RSCA Fellowship grant through SJSU. 

\bibliographystyle{IEEEtran}
\bibliography{aaai24}

\section{Appendix - PML Derivation}
Step 1. Given the split field form of the acoustic wave equation with pressure field $u: \mathbb{R}^2 \rightarrow \mathbb{R}^1$, velocity component in the $x$ direction $v_x: \mathbb{R}^2 \rightarrow \mathbb{R}^1$, and velocity component in the $y$ direction $v_y: \mathbb{R}^2 \rightarrow \mathbb{R}^1$, and speed of sound $c = \sqrt{a b}$.

\begin{align}
    \frac{\partial u}{\partial t} &= b \frac{\partial v_x}{\partial x} + b \frac{\partial v_y}{\partial y} \label{eq:u} \\
    \frac{\partial v_x}{\partial t} &= a \frac{\partial u}{\partial x} \label{eq:v_x} \\
    \frac{\partial v_y}{\partial t} &= a \frac{\partial u}{\partial y} \label{eq:v_y}
\end{align}

Step 2. The transformations are described by complex coordinate stretching in the frequency domain. The terms $\sigma_x$ and $\sigma_y$ are scalar functions which describe where the PML region is active and $\omega$ is frequency.

\begin{align}
    \frac{\partial}{\partial x} \rightarrow \frac{1}{1 + i \frac{\sigma_x}{\omega}}\frac{\partial}{\partial x} \label{eq:pml_x}\\
    \frac{\partial}{\partial y} \rightarrow \frac{1}{1 + i \frac{\sigma_y}{\omega}}\frac{\partial}{\partial y} \label{eq:pml_y}
\end{align}

Step 3. Convert \eqref{eq:u} to frequency domain and apply \eqref{eq:pml_x} and then \eqref{eq:pml_y}.

\begin{multline}
    -i\omega u = b\frac{\partial v_x}{\partial x} + b\frac{\partial v_y}{\partial y} + b\frac{\partial v_x}{\partial x}\frac{i\sigma_y}{\omega} + b\frac{\partial v_y}{\partial y}\frac{i\sigma_x}{\omega} \\ - u(\sigma_x + \sigma_y) - i\frac{u\sigma_x\sigma_y}{\omega} 
\end{multline}\label{eq:u_pml_frequency_domain}

Step 4. Every $\frac{i}{\omega}$ in \eqref{eq:u_pml_frequency_domain} is converted to a field in the frequency domain.

\begin{align*}
    -i\omega\Psi_y &= b\sigma_y\frac{\partial v_x}{\partial x} \\
    -i\omega\Psi_x &= b\sigma_x\frac{\partial v_y}{\partial y} \\
    -i\omega\Omega &= u\sigma_x\sigma_y \\
    -i\omega u &= b\frac{\partial v_x}{\partial x} + b \frac{\partial v_y}{\partial y} - u(\sigma_x + \sigma_y) + \Psi_x + \Psi_y - \gamma
\end{align*}

Step 5. Convert the system of equations back into the time domain.

\begin{align}
    \frac{\partial u}{\partial t} &= b \frac{\partial v_x}{\partial x} + b \frac{\partial v_y}{\partial y} - u(\sigma_x + \sigma_y) + \Psi_x + \Psi_y - \gamma \\
    \frac{\partial \Psi_x}{\partial t} &= b\sigma_x \frac{\partial v_y}{\partial y} \\
    \frac{\partial \Psi_y}{\partial t} &= b\sigma_y\frac{\partial v_x}{\partial x} \\
    \frac{\partial\gamma}{\partial t} &= u\sigma_x\sigma_y
\end{align}

Step 6. Convert \eqref{eq:v_x} to frequency domain and apply \eqref{eq:pml_x}.

\begin{equation}
    -i\omega v_x + v_x\sigma_x = a \frac{\partial u}{\partial x}
    \label{eq:v_x_pml_frequency_domain}
\end{equation}

Step 7. Convert \eqref{eq:v_y} to frequency domain and apply \eqref{eq:pml_x}.

\begin{equation}
    -i\omega v_y + v_y\sigma_y = a\frac{\partial u}{\partial y}
    \label{eq:v_y_pml_frequency_domain}
\end{equation}

Step 8. Convert \eqref{eq:v_x_pml_frequency_domain} and \eqref{eq:v_y_pml_frequency_domain} to time domain.

\begin{align}
    \frac{\partial v_x}{\partial t} &= a\frac{\partial u}{\partial x} - v_x\sigma_x \label{eq:v_x_pml_time_domain} \\
    \frac{\partial v_y}{\partial t} &= a\frac{\partial u}{\partial y} - v_y\sigma_y \label{eq:v_y_pml_time_domain}
\end{align}

The final system of equations used to simulate acoustic waves in 2D with a pml is given by the equations in Step 5 and Step 8.


\end{document}